\documentclass{PoS}
\usepackage[numbers]{natbib}
\usepackage{aas_macros}

\usepackage{multicol}

\title{Finding AGN with wide-field VLBI observations}

\ShortTitle{Finding AGN with wide-field VLBI observations}

\author{\speaker{Enno Middelberg}\\
        Astronomisches Institut der Ruhr-Universit\"at Bochum, Universit\"atsstr. 150, 44801 Bochum, Germany\\
        E-mail: \email{middelberg@astro.rub.de}}

\author{Adam Deller\\
        National Radio Astronomy Observatory, PO Box 0, Socorro, NM, 87801, USA\\
        E-mail: \email{adeller@nrao.edu}}

\author{John Morgan\\
        Curtin University of Technology, GPO BOX U1987, Perth, WA 6845, Australia\\
        E-mail: \email{john.morgan@curtin.edu.au}}

\author{Helge Rottmann\\
        Max-Planck-Institut f\"ur Radioastronomie, Auf dem H\"ugel 69, 53121 Bonn, Germany\\
        E-mail: \email{rottmann@mpifr-bonn.mpg.de}}

\author{Walter Alef\\
        Max-Planck-Institut f\"ur Radioastronomie, Auf dem H\"ugel 69, 53121 Bonn, Germany\\
        E-mail: \email{alef@mpifr-bonn.mpg.de}}

\author{Steven Tingay\\
        Curtin University of Technology, GPO BOX U1987, Perth, WA 6845, Australia\\
        E-mail: \email{steven.tingay@icrar.org}}

\author{Ray Norris\\
        Australia Telescope National Facility, PO Box 76, Epping NSW 1710, Australia\\
        E-mail: \email{ray.norris@csiro.au}}

\author{Uwe Bach\\
        Max-Planck-Institut f\"ur Radioastronomie, Auf dem H\"ugel 69, 53121 Bonn, Germany\\
        E-mail: \email{bach@mpifr-bonn.mpg.de}}

\author{Walter Brisken\\
        National Radio Astronomy Observatory, PO Box 0, Socorro, NM, 87801, USA\\
        E-mail: \email{wbrisken@nrao.edu}}

\author{Emil Lenc\\
        Australia Telescope National Facility, PO Box 76, Epping NSW 1710, Australia\\
        E-mail: \email{emil.lenc@csiro.au}}

\abstract{VLBI observations are a reliable method to identify AGN,
  since they require high brightness temperatures for a detection to
  be made. However, because of the tiny fields of view it is
  unpractical to carry out VLBI observations of many sources using
  conventional methods. We used an extension of the DiFX software
  correlator to image with high sensitivity 96 sources in the Chandra
  Deep Field South, using only 9h of observing time with the VLBA. We
  detected 20 sources, 8 of which had not been identified as AGN at
  any other wavelength, despite the comprehensive coverage of this
  field. The lack of X-ray counterparts to 1/3 of the VLBI-detected
  sources, despite the sensitivity of co-located X-ray data,
  demonstrates that X-ray observations cannot be solely relied upon
  when searching for AGN activity. Surprisingly, we find that sources
  classified as type 1 QSOs using X-ray data are always detected, in
  contrast to the 10\,\% radio-loud objects which are found in
  optically-selected QSOs. We present the continuation of this project
  with the goal to image 1450 sources in the Lockman Hole/XMM region.}

\FullConference{10th European VLBI Network Symposium and EVN Users Meeting: VLBI and the new generation of radio arrays\\
		September 20-24, 2010\\
		Manchester Uk}

\begin{document}

\section{Introduction}

Extragalactic radio surveys are an increasingly popular method to
investigate the evolution of star-forming and active
galaxies. However, their resolution is of the order of arcseconds and
hence not sufficient to separate the two emission processes. VLBI
observations are only sensitive to brightness temperatures of around
$10^6$\,K and can hence unambiguously identify AGN, provided the
inferred luminosity exceeds $2\times 10^{21}$\,W\,Hz$^{-1}$
[\citenum{Kewley2000}]. The high resolution and fringe rates of VLBI
impose severe logistical constraints on wide-field imaging, however,
and typical VLBI fields of view at cm wavelengths are around 10
arcseconds in diameter and hence unsuitable for surveying large
numbers of objects. Here we report on the first use of a new
multi-phase centre mode for the software correlator DiFX
[\citenum{Deller2007}], which allows one to produce normal-sized VLBI
data sets for a set of known target locations, and hence to image
hundreds of objects with a single observing run.

\section{The multi-phase centre mode of DiFX}

The challenge in wide-field VLBI observations is to overcome the
effects of time and bandwidth averaging, which reduce the amplitude of
sources away from the phase centre. The obvious solution is to
increase the temporal and spectral resolution at the correlation
stage, but this results in TB-sized visibility data sets, which are
difficult to process even on large computers. Whilst this approach has
recently been used by Wucknitz et al. and Morgan et al. (these
proceedings), we have developed an extension to the DiFX software
correlator [\citenum{Deller2010}], which processes multiple phase
centres in a single pass.

In this mode, the correlation is initially performed with high
frequency resolution which is sufficient to minimise bandwidth
smearing. Periodically, but still frequently enough to minimise time
smearing, the phase centre of the correlated data is shifted from its
initial location (which is usually the pointing centre) to a target
source location. This shift requires rotating the visibility phases of
each baseline by an amount equal to the difference in the geometric
delay between the final and initial source directions, multiplied by
the sky frequency. In effect, this corrects for the ``unapplied"
differential fringe rotation between the final and initial source
directions. This phase shift is repeated for each desired source
direction. After the phase shift is applied, the visibilities are
averaged in frequency and continue to be averaged in time. Eventually,
this results in an array of ``normal--sized" visibility datasets, with
one dataset per target source. The field of view of each of these
datasets is of the order of $13''$, at which point bandwidth and time
smearing would reduce the observed amplitudes by 5\,\%.

These data sets can be calibrated using standard methods. Furthermore,
since the phase response of a parabolic telescope is constant within
the primary beam, and since the geometric delays have been taken care
of at the correlation stage, the amplitude, phase, and delay
corrections can simply be copied from one data set to another.

\section{Observations and calibration}

On 3 July 2007 we have observed with the Very Long Baseline Array
(VLBA) at 1.4\,GHz a single pointing centred on the original CDFS
[\citenum{Luo2008}]. We used 8 dual-polarisation IFs with 8\,MHz
bandwidth each and two-bit sampling, resulting in a recording bitrate
of 512\,Mbps, which was the highest possible bitrate of the VLBA at
that time. The low declination of $-27^\circ$ allowed us to observe
the target for 9\,h only, yielding a total of 178 baseline-hours from
which we expect image sensitivities of 50\,$\mu$Jy (note that the low
elevations caused the system temperatures to be significantly higher
than at zenith during substantial parts of the observations). A
six-minute phase reference cycle time was used, with 5\,min allocated
to the target and 1\,min to the nearby strong (600\,mJy) calibrator
source NVSS\,J034838$-$274914. The data were recorded on disk and
shipped to the Max-Planck-Institut f\"ur Radioastronomie for
correlation.

Correlation was carried out on a 22-node computer cluster with 176
compute cores using a development version of the DiFX software
correlator. From previous ATCA observations [\citenum{Norris2006a}]
the positions of 96 sources within the VLBA's primary beam were known,
and so 96 data sets were produced using the new multi-phase centre
mode of DiFX. Calibration followed the standard steps for a
phase-referencing VLBI observation. After phase-referencing, the
target S503 was found to be bright enough for self-calibration, which
improved the coherence of the data. The S503 calibration tables were
then copied to the other data sets and all sources were imaged.

Unlike in compact-array interferometry, where the primary beam
correction is carried out in the image plane, we have corrected for
primary beam attenuation by calculating visibility gains. This is
possible only because the fields of view are very small in our
observations and so the attenuation due to the primary beam does not
vary significantly across the images. We have calculated the primary
beam attenuation at each station as a function of time and frequency
(the primary beam size changes with frequency, and since the VLBA
antennas have feeds offset from the optical axis they suffer from beam
squint, which also scales with frequency).

\section{Results}

All sources were imaged using natural and uniform weighting (yielding
restoring beams of 28.6$\times$9.3\,mas$^2$ and
20.8$\times$5.9\,mas$^2$, respectively), and using a 30\,\% Gaussian
taper at a $(u,v)$ distance of 10\,M$\lambda$, which resulted in a
restoring beam of 55.5$\times$22.1\,mas$^2$.

Out of the 96 targets, 20 were detected reliably, and one more target
was tentatively detected. We show here contour plots of three detected
targets as an illustration of the image quality.

\begin{figure}
\includegraphics[width=0.3\linewidth]{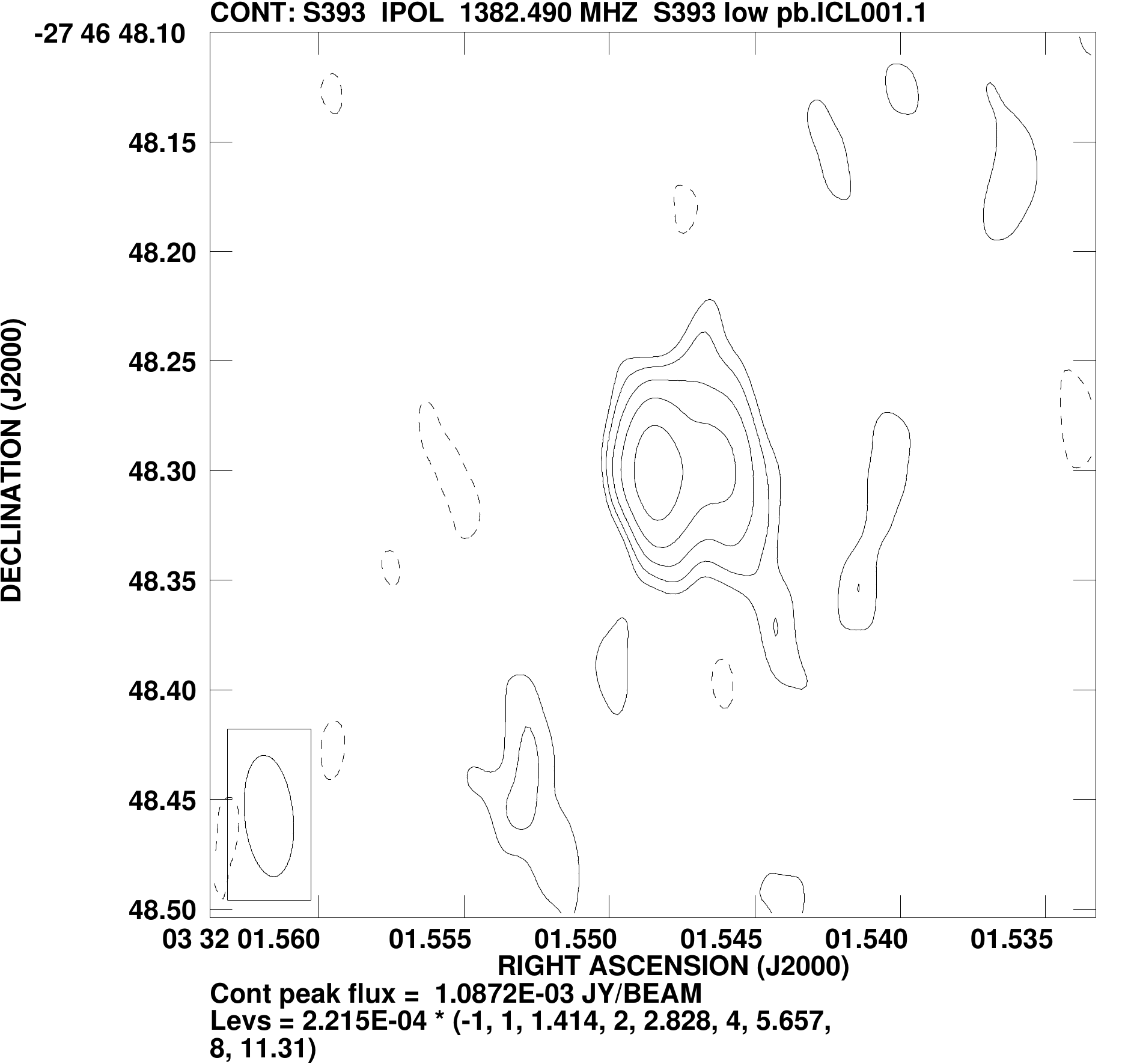} 
\includegraphics[width=0.3\linewidth]{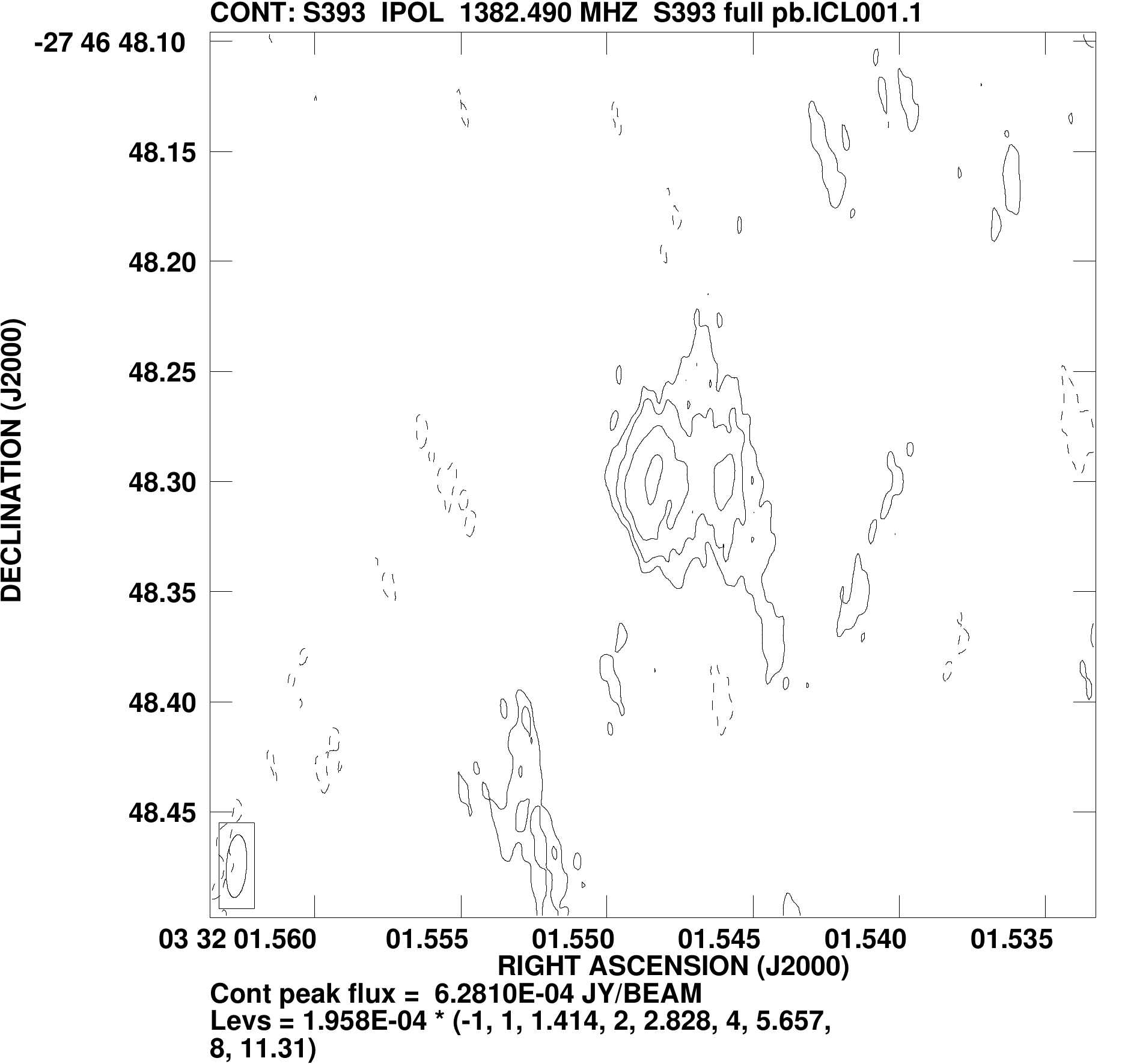}
\includegraphics[width=0.3\linewidth]{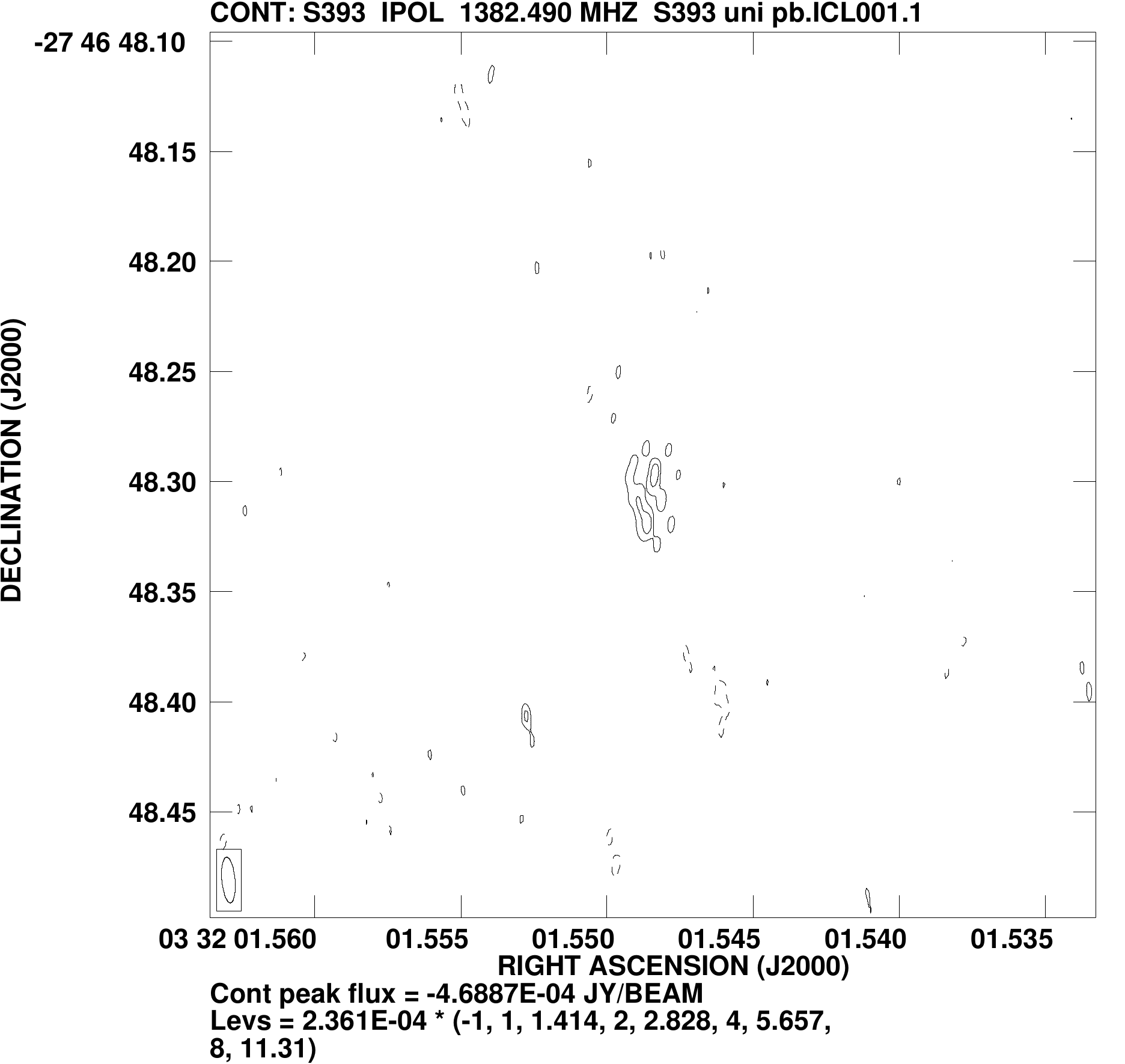} \\
\includegraphics[width=0.3\linewidth]{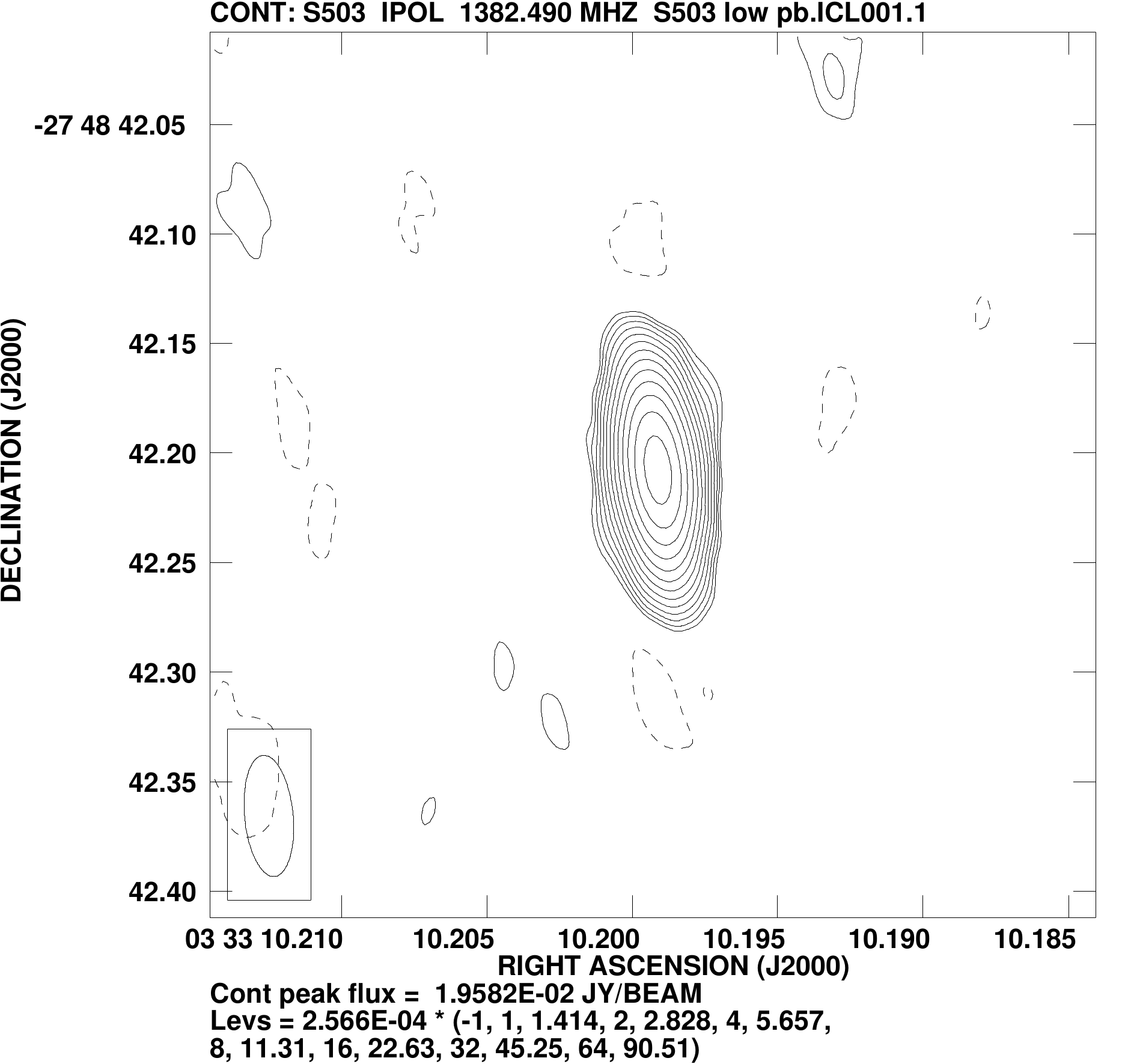} 
\includegraphics[width=0.3\linewidth]{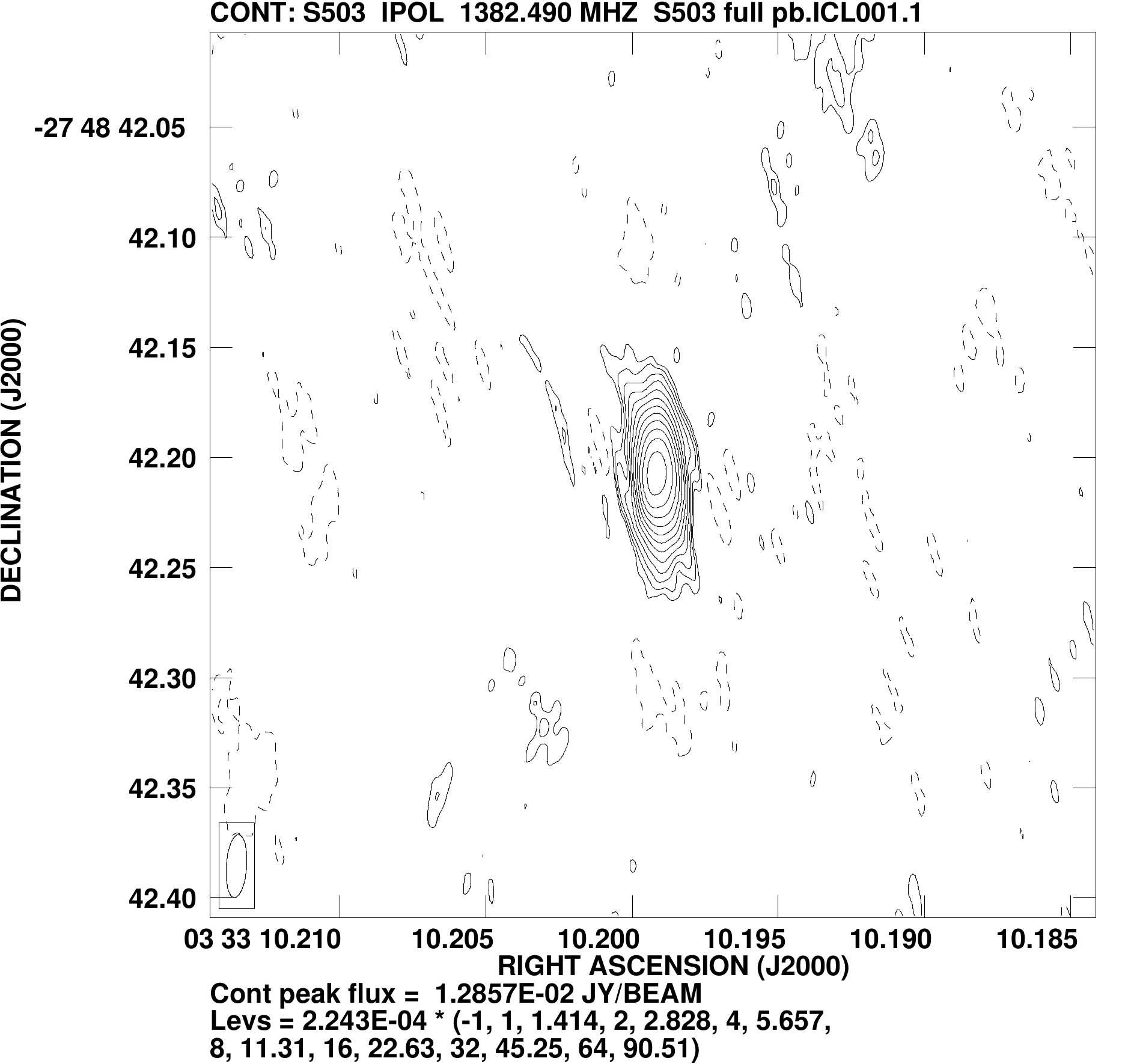}
\includegraphics[width=0.3\linewidth]{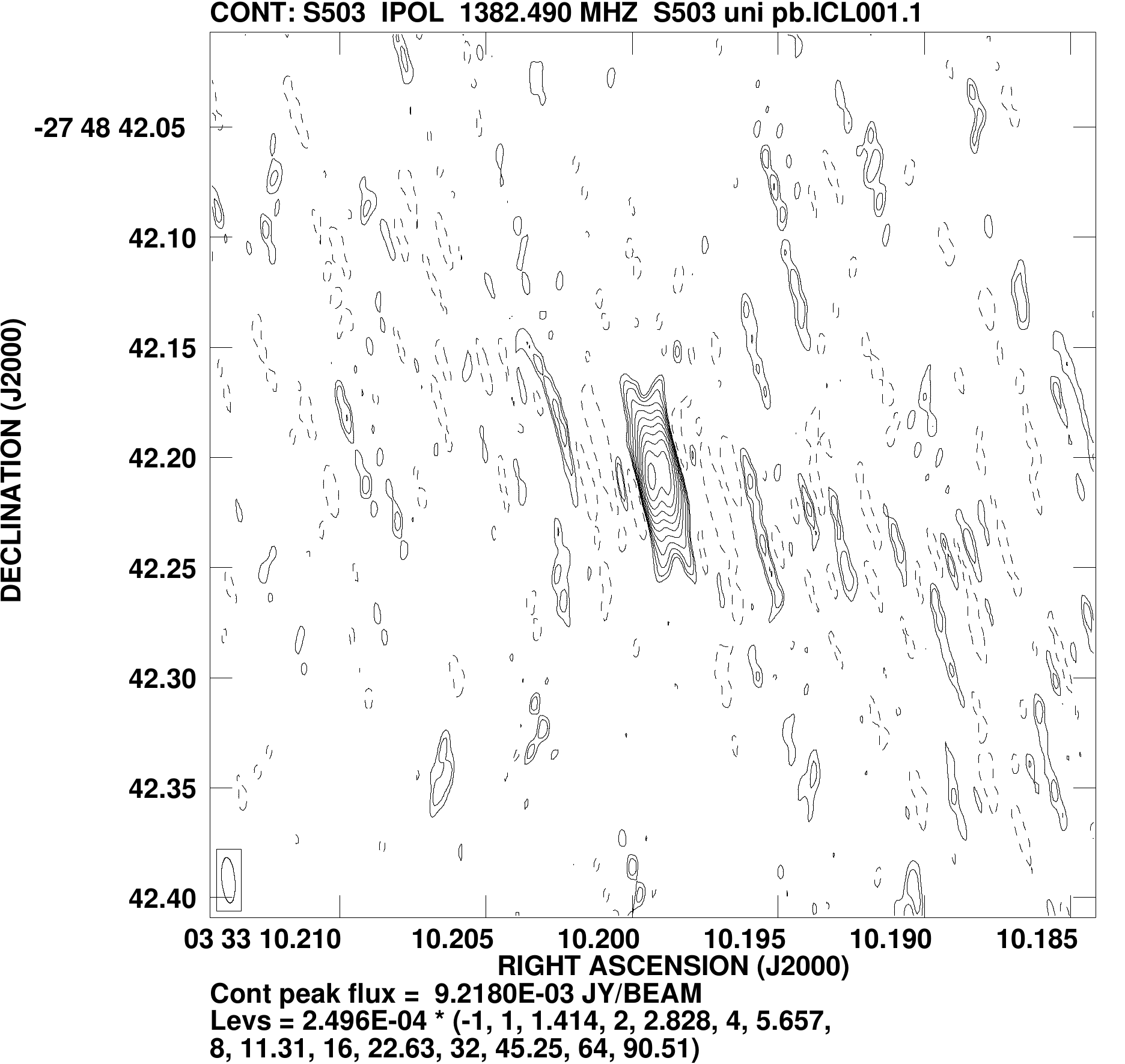} \\
\includegraphics[width=0.3\linewidth]{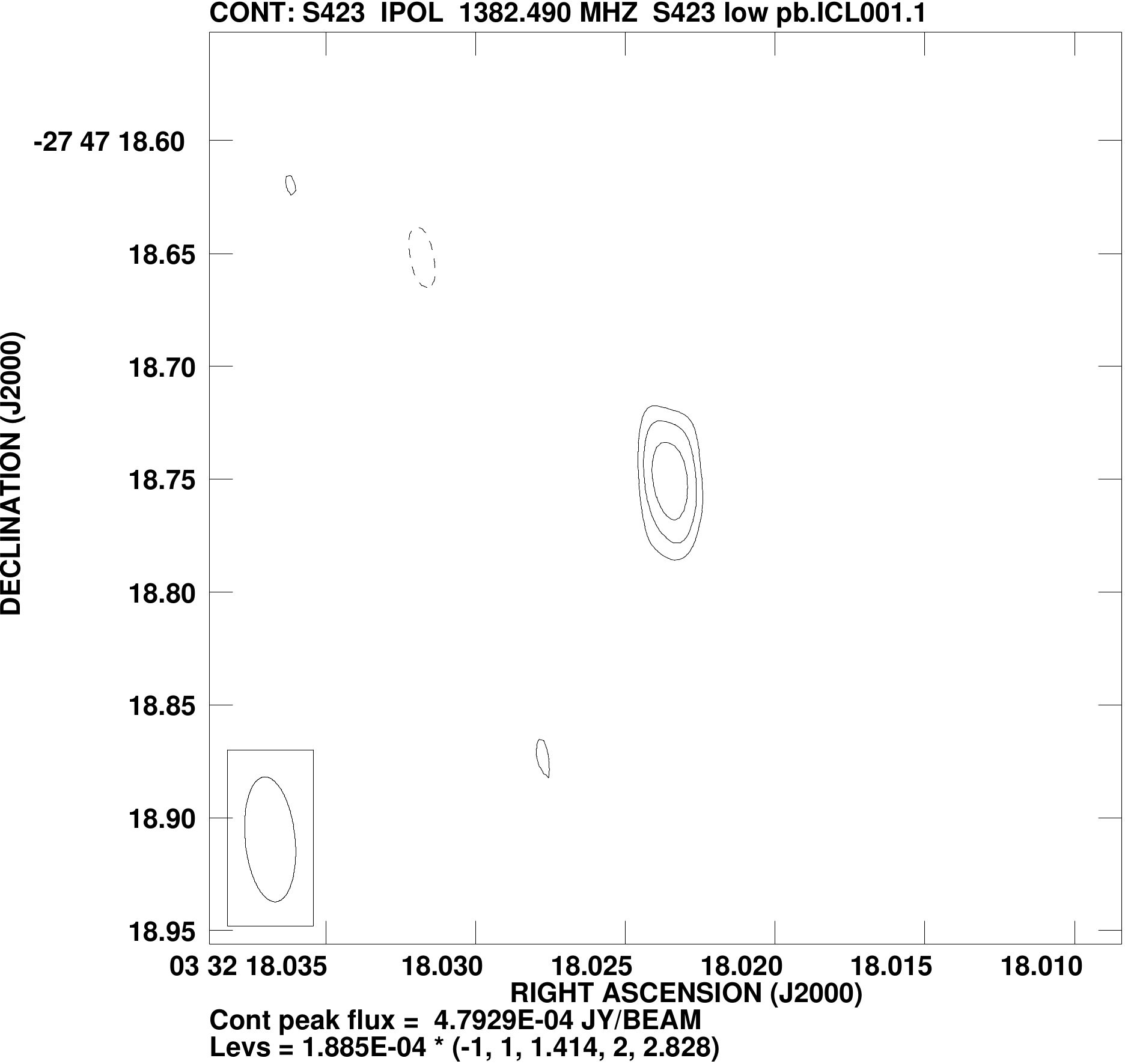} 
\includegraphics[width=0.3\linewidth]{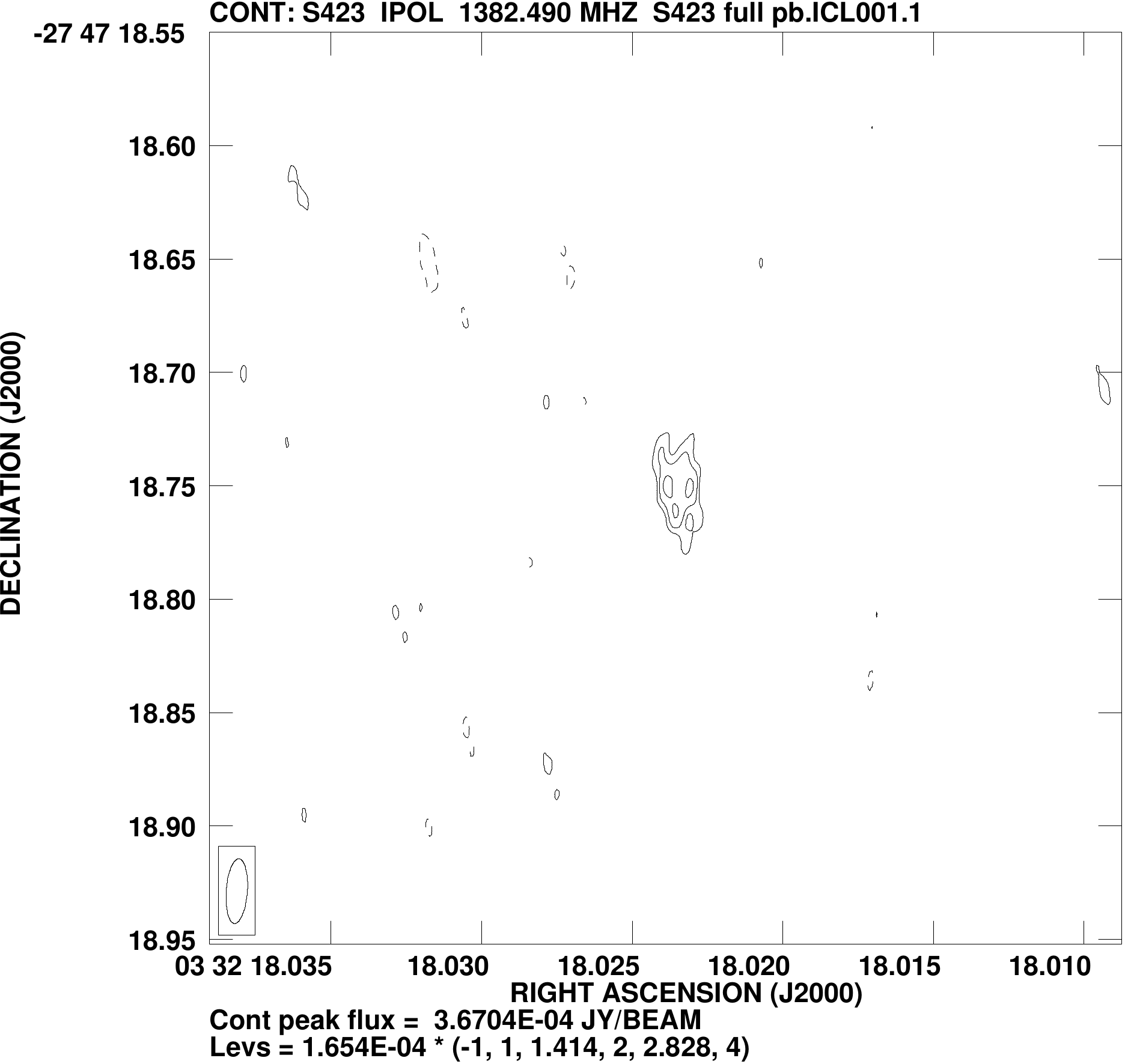}
\includegraphics[width=0.3\linewidth]{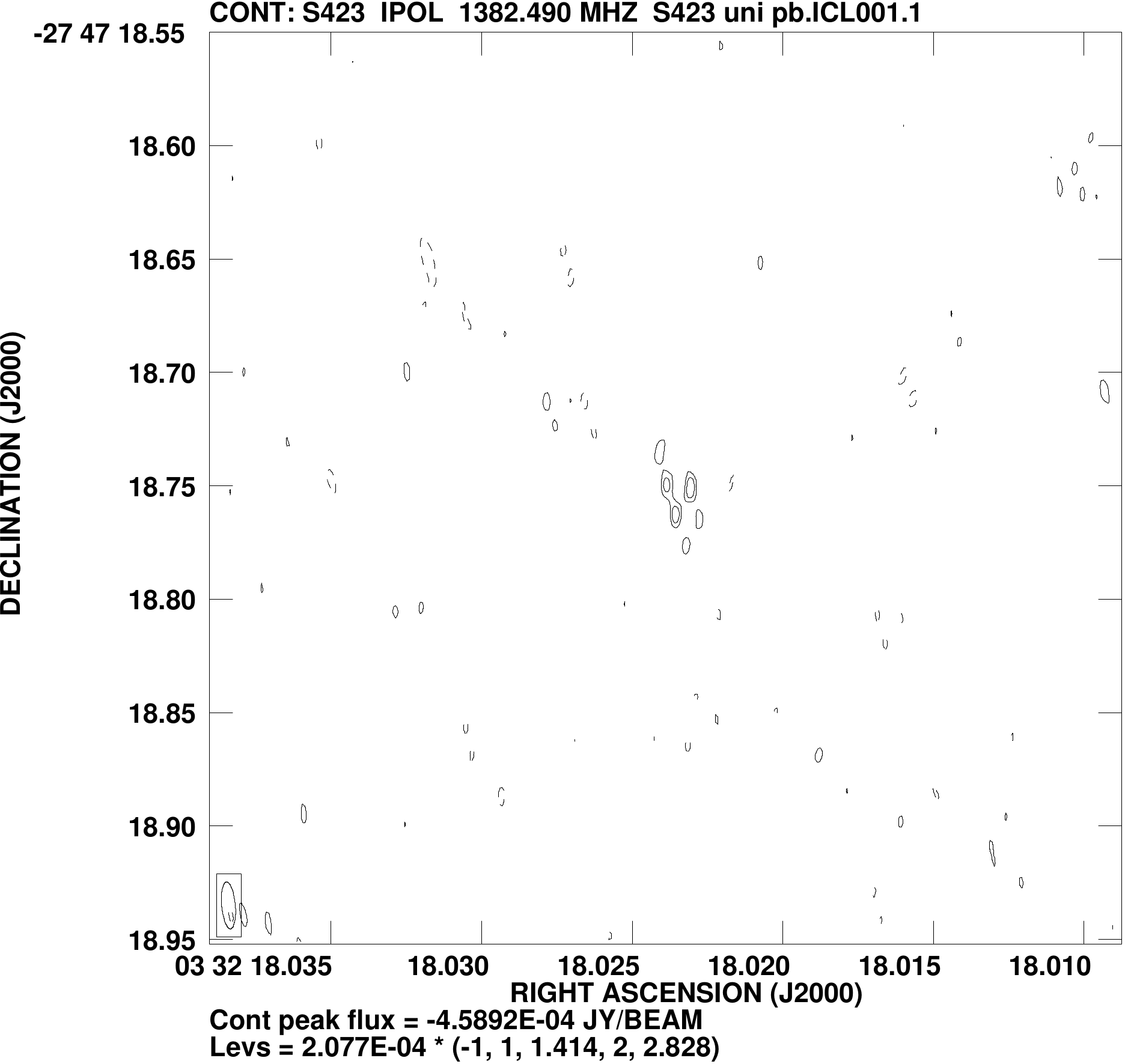} \\
\caption{Contour plots of three of the detected sources. Three images
  per source are shown in a row. {\it Left column:} naturally-weighted
  images; {\it middle column:} untapered, naturally-weighted image;
  {\it right column:} uniformly-weighted image. Positive contours
  start at three times the rms level of the images and increase by
  factors of $\sqrt{2}$. One negative contour is shown at three times
  the rms.}
\label{fig:images}
\end{figure}

We have cross-identified the 96 targets with the deep X-ray data in
the area [\citenum{Lehmer2005}, \citenum{Luo2008}], to obtain as much
evidence for AGN activity as possible. X-ray observations are claimed
to be a very direct tracer of AGN activity [e.g.,
  \citenum{Mushotzky2004}]. In the standard picture of AGN, the X-ray
emission originates very close to the supermassive black hole, is
relatively unabsorbed at energies of a few keV, and there is little
contamination from other sources, such as stars. Where redshifts were
available, we used the criteria by [\citenum{Szokoly2004}] to
determine an object's nature using the X-ray data alone.

Surprisingly, we have detected with the VLBA 7 sources which have no
X-ray counterparts. One of these sources is in a region covered by a
2\,Ms Chandra exposure [\citenum{Luo2008}]. All these targets have
redshifts exceeding 0.1, hence their radio luminosities suggest that
they are AGN and not starbursts [\citenum{Kewley2000}].

We also investigated the detection rate of those VLBI-detected targets
with available X-ray counterparts, as a function of X-ray source
type. We found that all sources classified as type 1 QSOs by the
criteria by [\citenum{Szokoly2004}] were detected (there were no type
2 QSOs in our sample). This is in stark contrast to optically selected
QSOs, of which typically only a fraction of 10\,\% is radio-loud.

We investigated potential differences between VLBI-detected and
undetected sources as a function of spectral index (using the data by
[\citenum{Kellermann2008}]) and redshift, but found no significant
differences.

One detected source in our sample, S423, shows no indication of AGN
activity in optical and X-ray observations. However, it is clearly
detected, and given its redshift of 0.73 it must contain an
AGN. Another object, S443, has been tentatively detected. However, the
VLBI core is offset from the centre of the galaxy and can therefore
not be taken as evidence for an AGN. We note that the radio luminosity
of this source, $5\times 10^{21}$\,W\,Hz$^{-1}$, is similar to one of
the brightest radio supernovae, SN1986J. We therefore consider it
possible that S443 is a radio supernova.

\section{Conclusions and outlook}

This pilot project can be regarded as a success, and the multi-phase
centre mode is now publicly available at the VLBA. We have therefore
embarked on a project to carry wide-field VLBI even further, using
mosaicing of separate pointings, to increase the sensitivity of the
observations over wide areas.

We have observed with the VLBA the three pointings observed by
[\citenum{Ibar2009}] with the VLA, resulting in an on-axis sensitivity
of 23\,uJy in each pointing. Around 330 phase centres were correlated
in each run, targeting only those 508 sources with a 1.4\,GHz flux
density of more than 100\,$\mu$Jy. The observations resulted in a
total of 1\,TB of visibility data. In a follow-up project we will
attempt to image all 1450 sources with a sensitivity on par with the
VLA observations, when the VLBA recording rate has been increased to
at least 2\,Gbps.

\begin{figure}
\includegraphics[width=0.45\linewidth]{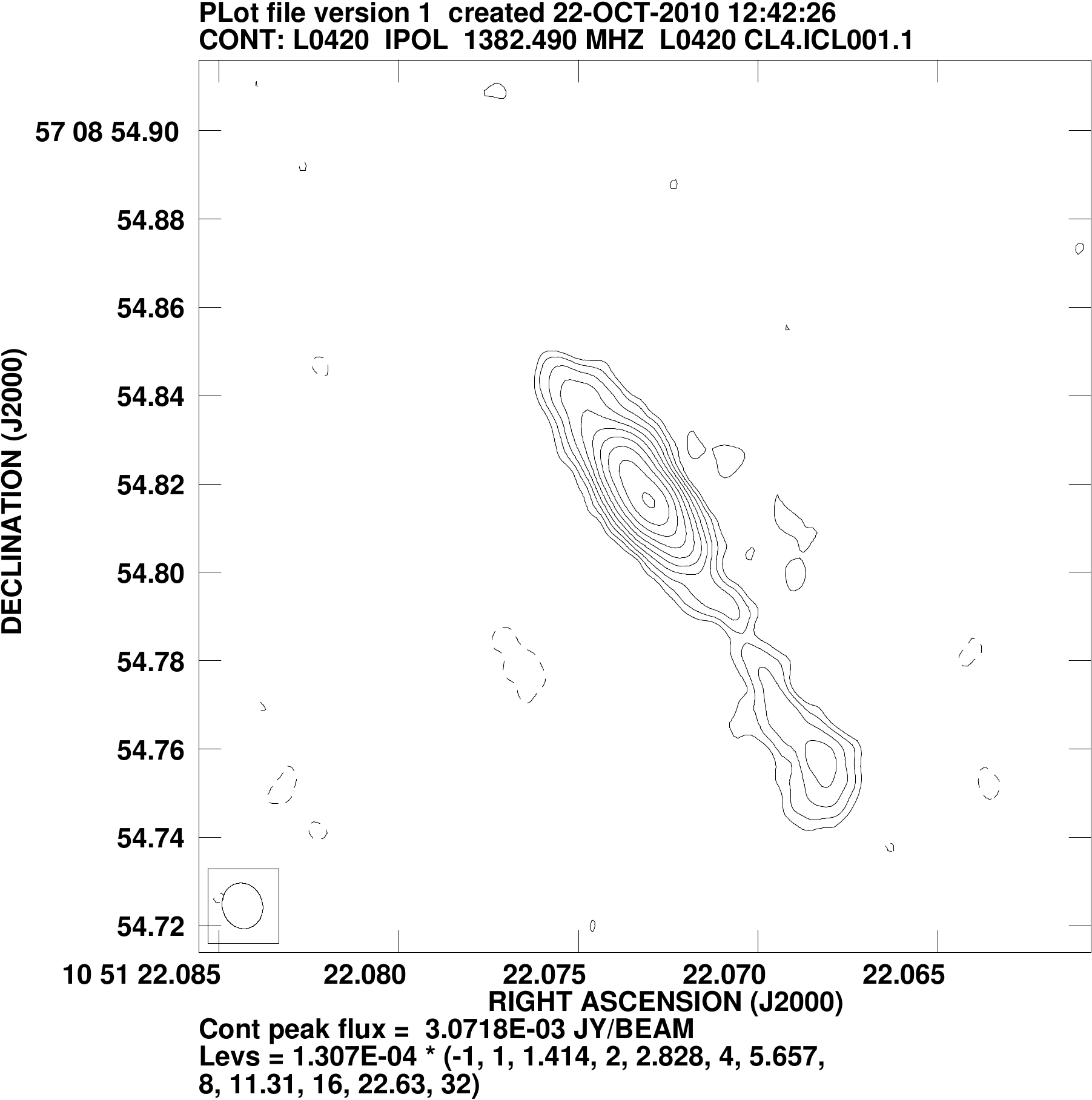} 
\includegraphics[width=0.45\linewidth]{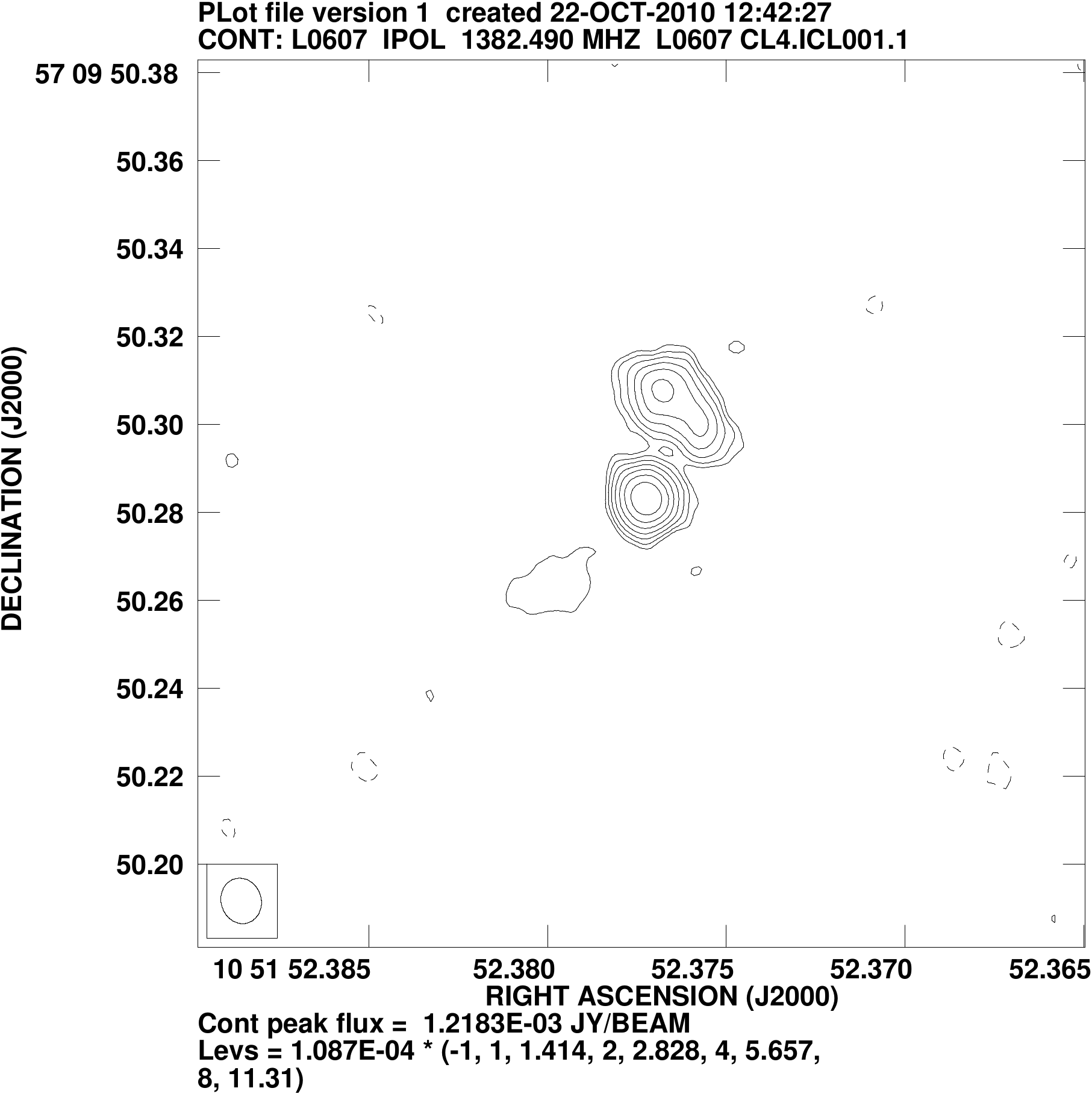}\\
\caption{Contour plots of two targets detected in the Lockman Hole
  East. Contours are drawn at three times the rms level, and one
  negative contour is shown.}
\label{fig:images2}
\end{figure}

The calibration of the centre pointing has been finished, but the
images have not yet been analysed. We show in Fig.~\ref{fig:images2}
two example images of interestingly-looking objects. The image
fidelity is higher than that achieved with the CDFS observations,
something we attribute to the higher elevation, and consequently
better uv-coverage and calibration transfer, of the Lockman Hole East
field.

\begin{multicols}{2}

\end{multicols}

\end{document}